\documentclass[preprint,12pt]{elsarticle}

\usepackage{comment}
\usepackage{amssymb}
\usepackage{amsmath}
\usepackage{hyperref}   
\usepackage{url}        

\journal{Physica A: Statistical Mechanics and its Applications}

\begin{document}

\begin{frontmatter}



\title{Diffusion Signals Reveal Hidden Connections: A Physics-Inspired Framework for Link Prediction via Personalized PageRank Signals}


\author[label1]{Huilin Wang}

\author[label2]{Wenjun Zhang}
\ead{wenjun@ahtcm.edu.cn}

\author[label1]{Weibing Deng}
\ead{Wdeng@mail.ccnu.edu.cn}

\affiliation[label1]{%
  organization={Key Laboratory of Quark and Lepton Physics (MOE) and Institute of Particle Physics, Central China Normal University},
  addressline={Wuhan 430079},
  city={Wuhan},
  postcode={430079},
  country={China}
}
\affiliation[label2]{%
  organization={School of Information Engineering, Anhui  University of Chinese Medicine},
  addressline={Hefei 230012},
  city={Hefei},
  postcode={230012},
  country={China}
}

\begin{abstract}

Link prediction in complex networks--identifying the missing or future connections--remains a cornerstone problem for understanding network evolution and function, yet existing methods struggle to balance computational efficiency with theoretical rigor across heterogeneous topologies. This work introduces a physically principled framework, Diffusion Distance with Personalized PageRank (D-PPR), which unifies static topology with dynamic information flow by modeling nodes as signal sources propagating through the network via Personalized PageRank (PPR) vectors. The method quantifies node-pair similarity through the graph Laplacian-governed diffusion distance between their topology-aware signal distributions, thereby bridging microscopic interactions with macroscopic network dynamics. Systematic benchmarking on synthetic (Barabási-Albert, LFR) and seven large-scale real-world networks spanning technology, biology, and social domains demonstrates that D-PPR achieves highly competitive performance, yielding favorable results when compared to representative local and global heuristics, particularly in sparse and modular networks. These findings establish a rigorous foundation for physics-inspired link prediction by revealing that incorporating dynamical processes into structural similarity metrics enables deeper insights into network connectivity patterns, offering both methodological advances and new theoretical perspectives on the interplay between topology and dynamics.

\end{abstract}

\begin{highlights}
    \item A physics-inspired link prediction framework based on network diffusion is proposed.
    \item Personalized PageRank vectors are leveraged as rich, multi-scale structural probes of nodes.
    \item The framework unifies static topology and network dynamics to achieve high robustness and accuracy.
\end{highlights}

\end{frontmatter}


\section{Introduction}

Complex networks have become a universal paradigm for modeling systems of interacting components, ranging from the intricate web of protein-protein interactions within cells\cite{rao2014protein,braun2012history} to the sprawling architecture of the Internet\cite{yuan2023structural,faloutsos1999power}. A fundamental challenge in this domain lies in uncovering the governing principles that shape network structure and evolution\cite{stam2014modern,newman2003structure,albert2002statistical}. The link prediction \cite{lu2011link,liben2003link,martinez2016survey}problem—which seeks to infer missing\cite{zhou2009predicting} or future connections \cite{lu2015toward}based on a network's current topology—serves not only as a practical engineering task\cite{su2020link,mccoy2021biomedical} but also as a powerful theoretical tool for probing these fundamental organizational principles.

Existing approaches to link prediction can be broadly classified into two primary families based on the scope of topological information they utilize: local and global methods \cite{lu2010link}. Local methods, such as the widely-used Common Neighbors (CN) index and the Adamic-Adar (AA) index\cite{adamic2003friends}, are computationally efficient as they only consider the immediate neighborhood of a node pair. However, this reliance on nearest-neighbor information means they inherently overlook the rich, long-range correlations that characterize most real-world networks. In contrast, global methods, exemplified by the Katz Index\cite{katz1953new} and the Leicht-Holme-Newman (LHN) index\cite{leicht2006vertex}, consider the entire network topology, often by summing over all possible paths between nodes. Although this comprehensive view can yield higher accuracy, it typically comes at the cost of prohibitive computational complexity, making such methods impractical for very large networks. This fundamental trade-off between efficiency and accuracy motivates the search for novel frameworks that can bridge this gap.

These limitations suggest that a deeper understanding of node similarity\cite{lu2009similarity} may arise not from static path enumeration but from the dynamical processes unfolding on the network. A network's topology inherently constrains the flow of information, influence, or physical quantities\cite{boccaletti2006complex}. The diffusion process, governed by the graph Laplacian \cite{merris1995survey}, provides a canonical framework for modeling such flows and naturally defines a notion of "effective distance" between nodes.

In this work, we leverage this insight to develop a novel theoretical framework for link prediction. We propose that the likelihood of a link between two nodes is positively correlated with their structural position, as quantified through a dynamical process that integrates diffusion distance and personalized PageRank. Specifically, we introduce the \textbf{Diffusion Distance with Personalized PageRank (D-PPR)} framework, which comprises two key components:

\begin{enumerate}
    \item \textbf{Personalized PageRank (PPR) as a structural probe}: Unlike simple local metrics, the PPR vector \cite{zhang2016approximate,park2019survey} captures the multi-scale influence field projected by a node across the entire network, reflecting both direct and indirect connections.
    \item \textbf{Diffusion distance as a similarity metric}: By applying the network diffusion process to the space of PPR signals, we define a ``diffusion distance'' that quantifies how closely two nodes' influence fields propagate. This provides a robust measure of their structural relationship.
\end{enumerate}

Through systematic numerical experiments on both synthetic (Barabási-Albert, LFR) and seven large-scale real-world networks spanning technology, biology, and social domains, we demonstrate that this dynamics-informed framework consistently outperforms traditional static topological measures. Our findings establish a rigorous foundation for physics-inspired link prediction by revealing that incorporating dynamical processes into structural similarity metrics enables deeper insights into network connectivity patterns, offering both methodological advances and novel theoretical perspectives on the interplay between topology and dynamics.

The remainder of this paper is structured as follows. Section 2 details the theoretical underpinnings of our framework, including the network diffusion process and Personalized PageRank as a structural probe. Section 3 formally presents the steps of our proposed D-PPR method. In Section 4, we conduct extensive numerical simulations on both synthetic and real-world networks to validate our approach against baseline methods. Finally, Section 5 discusses the implications of our findings and concludes the paper.

\section{Theoretical Framework}

\subsection{The Diffusion Process on Networks}
Consider an undirected and unweighted graph $G=(V, E)$, with $n=|V|$ nodes. Its topology is encoded in the adjacency matrix $A$ and the diagonal degree matrix $D$. The graph Laplacian, $L = D - A$, is a fundamental operator that can be viewed as a discrete analogue of the Laplace operator in continuous space. It governs the diffusion of a scalar quantity (e.g., heat) across the network, described by the differential equation:
\begin{equation}
\frac{d\mathbf{s}(t)}{dt} = -L\mathbf{s}(t)
\end{equation}
where $\mathbf{s}(t)$ is a vector representing the quantity at each node at time $t$. The solution is given by $\mathbf{s}(t) = e^{-Lt}\mathbf{s}(0)$, where $s(0)$ is the initial state. This process naturally defines a distance metric. As shown in\cite{segarra2015diffusion,zhang2018statistical} , the diffusion distance between two initial signals $s_u$ and $s_v$ is given by:
\begin{equation}
d_{\text{diff}}(\mathbf{s}_u, \mathbf{s}_v) = \left\| (I + \alpha L)^{-1}(\mathbf{s}_u - \mathbf{s}_v) \right\|_2
\label{eq:diffusion_distance}
\end{equation}
where $I$ is the identity matrix and $\alpha > 0$ is the diffusion coefficient. This metric quantifies the dissimilarity of the total diffused effects of the two signals.

\subsection{Personalized PageRank as a Structural Probe}
The effectiveness of the diffusion distance depends critically on the choice of the initial signals $\mathbf{s}_u$ and $\mathbf{s}_v$. A simple choice, like a one-hot vector, only captures the node's identity. To create a more informative signal, we use Personalized PageRank (PPR). The PPR vector for a source node $u$, denoted $\mathbf{s}_u$, is the stationary distribution of a random walker that teleports back to node $u$ with probability $1-\beta$ at each step. It is the solution to:
\begin{equation}
\mathbf{s}_u = (1-\beta) \mathbf{e}_u + \beta A D^{-1} \mathbf{s}_u
\end{equation}
where $\mathbf{e}_u$ is the one-hot vector for node $u$. The PPR vector can be interpreted as a node's ``influence field" or a measure of its proximity to all other nodes, making it an ideal candidate for a structural probe.

\subsection{D-PPR: A Unifying Framework}
Our D-PPR framework combines these two concepts to define a link score. For any pair of nodes $(u, v)$, the D-PPR distance is calculated by applying the diffusion distance metric (Eq. \ref{eq:diffusion_distance}) to their respective PPR signal vectors:
\begin{equation}
d_{\text{D-PPR}}(u, v) = \left\| (I + \alpha L)^{-1}(\mathbf{s}_u - \mathbf{s}_v) \right\|_2
\end{equation}
The final link prediction score is then defined as the reciprocal of this distance, reflecting the intuition that smaller distance implies higher similarity:
\begin{equation}
\text{Score}(u, v) = \frac{1}{d_{\text{D-PPR}}(u, v) + \epsilon}
\end{equation}
where $\epsilon$ is a small regularization constant. This score elegantly captures the structural relationship between $u$ and $v$ by measuring how similarly their global influence fields propagate through the network's topological fabric.

\subsection{Computational Complexity and Practical Considerations}

The superior performance of the D-PPR framework comes at a computational cost that is higher than that of simple local heuristics. Here, we analyze the complexity of our method and discuss its practical implementation. The overall computation for a pair of nodes $(u, v)$ involves two main steps: calculating the PPR vectors and computing the diffusion distance.

 \vspace{0.5em}\noindent{1 Personalized PageRank (PPR) Vector Computation}\vspace{0.5em}

The PPR vector $s_u$ for a node $u$ is the solution to the linear system $(I - \beta A D^{-1})\mathbf{s}_u = (1 - \beta)\mathbf{e}_u$, as implied by Equation~(3). While this can be solved directly, a more efficient approach for sparse networks is to use iterative methods such as the Power Method or topic-sensitive PageRank algorithms\cite{zhang2016approximate}. The complexity of computing a single PPR vector up to a desired precision is approximately $O(k \cdot m)$, where $m$ is the number of edges and $k$ is the number of iterations required for convergence. In a link prediction task, PPR vectors need to be computed for a set of target nodes, which constitutes the most computationally intensive part of the preprocessing phase.

\vspace{0.5em}\noindent{ 2 Diffusion Distance Calculation}\vspace{0.5em}

The calculation of the diffusion distance, $d_{\text{D-PPR}}(u, v) = \|(I + \alpha L)^{-1}(\mathbf{s}_u - \mathbf{s}_v) \|_2$, involves solving a linear system of the form $(I + \alpha L)x = b$, where $b = \mathbf{s}_u - \mathbf{s}_v$. A naive approach of directly inverting the matrix $(I + \alpha L)$ would have a prohibitive complexity of $O(n^3)$, where $n$ is the number of nodes.

However, since the graph Laplacian $L$ is sparse for most real-world networks, this system can be solved much more efficiently. We employ iterative numerical methods, such as the Conjugate Gradient (CG) method, which is well-suited for solving systems involving symmetric positive-definite matrices like $(I + \alpha L)$. The complexity of the CG method per iteration is proportional to the number of non-zero entries in the matrix, which is $O(m)$. The total complexity for computing the distance for one pair is therefore $O(k_{cg} \cdot m)$, where $k_{cg}$ is the number of CG iterations needed for convergence.

\vspace{0.5em}\noindent{3 Overall Perspective}\vspace{0.5em}

The D-PPR framework's complexity is primarily determined by the iterative computations of PPR vectors and the solution of the diffusion equation. While significantly more demanding than local indices like Common Neighbors \cite{newman2001clustering,yang2016predicting,yao2016link,li2018similarity} (whose cost is tied to node degrees), its complexity is comparable to other global methods like the Katz Index\cite{katz1953new,lu2009similarity,cocskun2021fast,mojarad2021development}, which also involves matrix inversion or equivalent iterative schemes.

This trade-off between computational cost and predictive power is central to our framework. The investment in computation allows D-PPR to capture rich, non-local topological and dynamical information, leading to the robust and superior performance demonstrated in our experiments. For very large-scale networks, the efficiency can be further improved by leveraging approximation algorithms for PPR and parallel computing architectures, which we leave as an avenue for future work.

\section{The Proposed D-PPR Method}

\begin{figure}[h!]
    \centering
    \includegraphics[width=\textwidth]{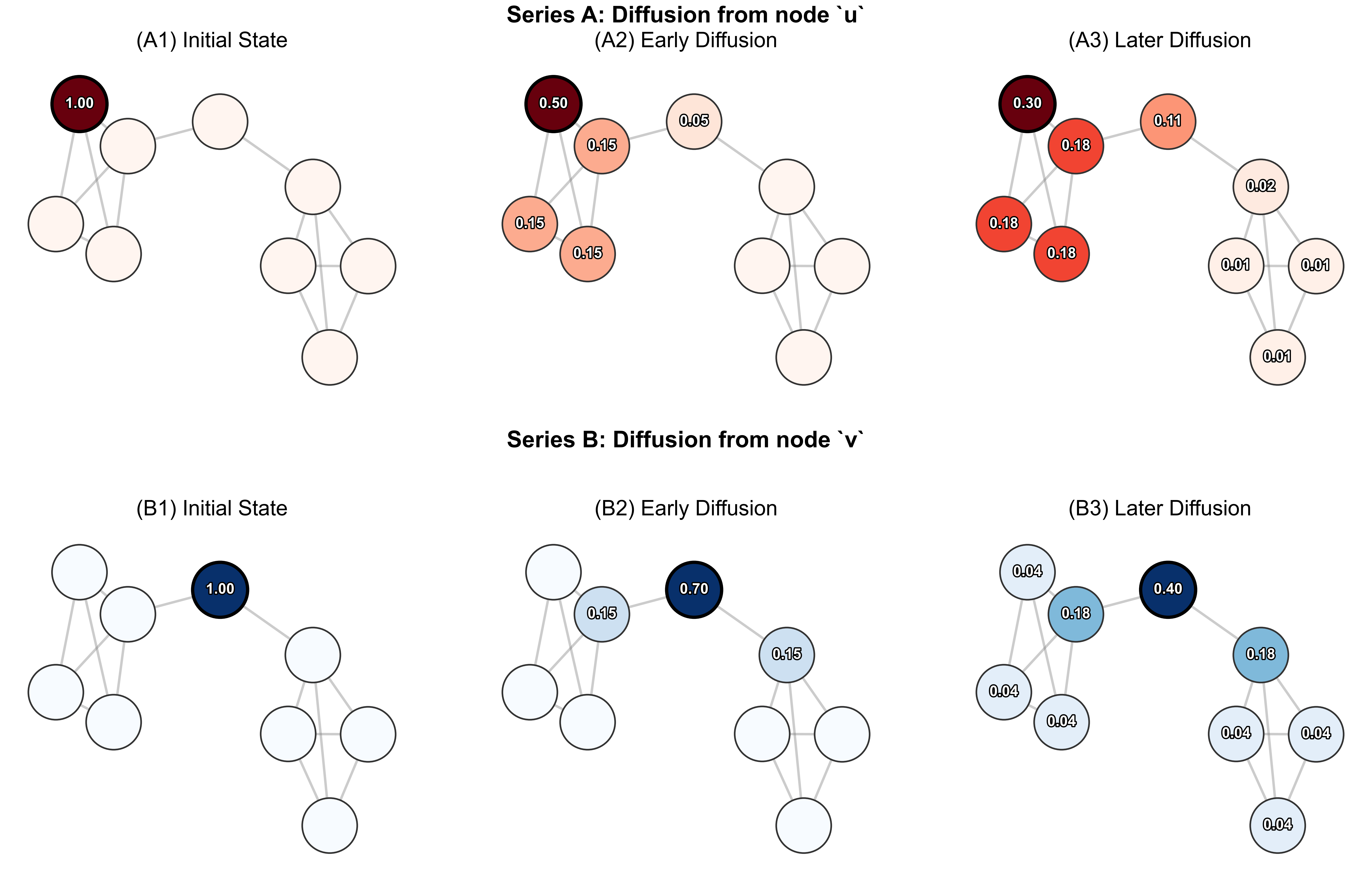}
    \caption{
    Visualization of the D-PPR Method's Core Intuition.
    The panels (Series A \& B) visualize signals diffusing from two distinct nodes over time. Our method quantifies the similarity of these diffusion patterns using a \textit{diffusion distance}. This distance, applied to Personalized PageRank signals, is then used to compute a final link score.
}
    \label{fig:conceptual_overview}
\end{figure}

Before delving into the mathematical details, we first present a visual intuition for our D-PPR framework in Figure~\ref{fig:conceptual_overview}. The core idea is to quantify the structural similarity between two nodes, say $u$ and $v$, by treating them as sources of unique signals that propagate across the network, akin to a physical diffusion process. Figure~\ref{fig:conceptual_overview} illustrates this concept by showing how signals originating from different nodes spread through the network topology over time.

The process, as illustrated in the figure, unfolds in four key steps:
\begin{enumerate}
    \item[\textbf{1}] We begin with a pair of nodes, $u$ and $v$, within the network topology.
    \item[\textbf{2}] For each node, we compute its Personalized PageRank (PPR) vector, denoted $\mathbf{\pi}_u$ and $\mathbf{\pi}_v$. This vector serves as a rich structural "fingerprint," capturing the node's multi-scale proximity to all other nodes. As shown in the bar plots, the PPR profile reveals how a node's influence is distributed throughout the network.
    \item[\textbf{3}] We then calculate the Diffusion Distance between these two signal vectors. This is achieved by taking the difference of the PPR vectors ($\mathbf{\pi}_u - \mathbf{\pi}_v$) and simulating its diffusion across the graph, a process governed by the graph Laplacian $L$. The norm of the resulting diffused vector gives us the distance $d(u,v)$.
    \item[\textbf{4}] Finally, the link prediction score $S(u,v)$ is defined as the reciprocal of this distance. A smaller distance signifies that the nodes are structurally more equivalent, resulting in a higher similarity score and a stronger prediction of a link.
\end{enumerate}

The following subsections will formalize each of these steps.

\section{Numerical Simulations and Results}

\subsection{Experimental Protocol}
To rigorously assess the performance and generalizability of our D-PPR framework, we design a comprehensive evaluation protocol. We test our method against a selection of representative baseline algorithms on a diverse suite of network datasets. The specific components of our experimental setup are detailed below.
\begin{itemize}
    \item Model Networks: We use two canonical generative models to systematically investigate the impact of specific network properties.
    \begin{itemize}
        \item For the Barabási-Albert (BA) model\cite{barabasi1999emergence,bertotti2019configuration}, we generate networks with a fixed number of nodes ($N=500$) and vary the attachment parameter $m$ (from $m=2$ to $m=8$) to create graphs with different average degrees, allowing us to precisely study the effect of network density.
        \item For the LFR benchmark model\cite{yang2017hierarchical,gancio2021community,lancichinetti2009community}, we generate networks with $N=250$ nodes and vary the mixing parameter $\mu$ from 0.1 to 0.7 to tune the modularity. Other LFR parameters were held constant to isolate the effect of community structure: the degree distribution exponent $\tau_1=3$, the community size exponent $\tau_2=1.5$, an average degree of 5, and a minimum community size of 20.
    \end{itemize}
    
    \item Real-World Networks: We analyze seven empirical networks from a range of domains to demonstrate the framework's practical applicability: Karate (social), Citation, Email (communication), Metabolic (biological), WWW(technological), and two transportation networks, Air-China and London-Tube. Key statistics are summarized in Table~\ref{tab:datasets}.
\end{itemize}
We compare D-PPR with three baseline methods: Common Neighbors (CN), Adamic-Adar (AA)\cite{adamic2003friends}, and the Katz Index. For evaluation, we withhold 10\% of the edges as a positive test set and sample an equal number of non-existent edges as a negative test set. Performance is measured using AUPR\cite{flach2015precision,pathak2017ensemble,zhou2023discriminating}.

\begin{table}[h!]
\centering
\caption{Statistics of the network datasets used in our numerical simulations.}
\label{tab:datasets}
\begin{tabular}{l r r r r}
\hline
\hline
Dataset & Nodes & Edges & Avg. Degree & Type \\
\hline
Karate Club      & 34             & 78              & 4.59                 & Social \\
Citation         & 449,673        & 4,685,458       & 20.84                & Collaboration \\
Email            & 57,194         & 92,442          & 3.23                 & Communication \\
Metabolic        & 1,039          & 4,741           & 9.13                 & Biological \\
WWW             & 325,729        & 1,090,108       & 6.69                 & Technological \\
Air-China        & 949            & 10,757          & 22.67                & Transportation \\ 
London-Tube      & 301            & 358             & 2.38                 & Transportation \\
\hline
\hline
\end{tabular}
\end{table}

\subsection{Simulations on Model Networks}
\begin{figure}[h!]
    \centering
    \includegraphics[width=0.8\textwidth]{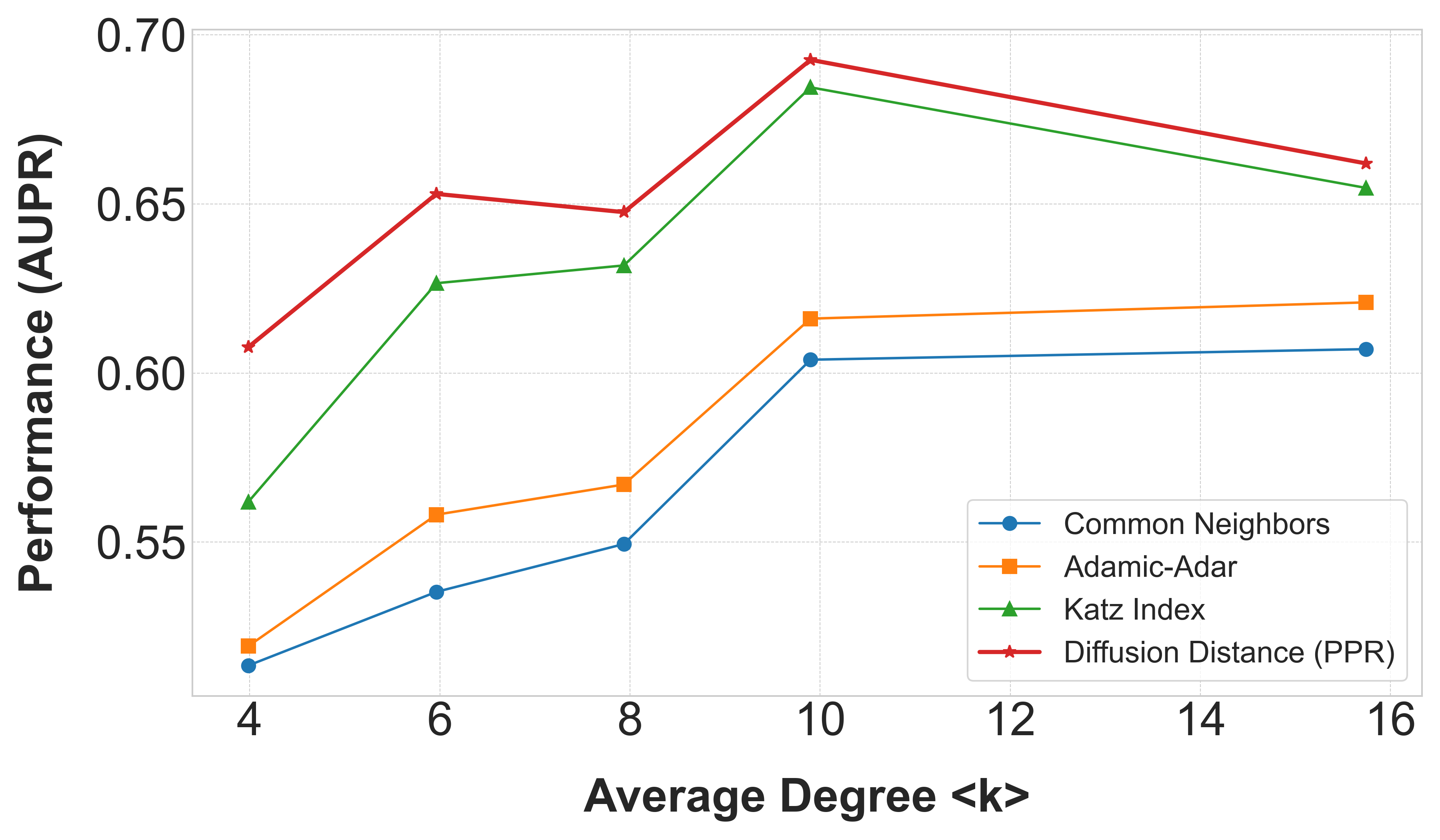}
    \caption{Performance of link prediction methods as a function of network density on Barabási-Albert (BA) model networks. The plot illustrates how the AUPR score for each method changes as the average degree $<k>$ of the network increases. Networks were synthetically generated using the BA model with $N=500$ nodes. The results indicate that the performance of global methods, particularly our proposed D-PPR, improves substantially with increasing density. This suggests that these methods effectively leverage the greater abundance of paths available in denser graphs.}
    \label{fig:ba_density}
\end{figure}

Effect of Density: Figure \ref{fig:ba_density} shows that on BA networks, the performance of global methods, including D-PPR, increases with network density. D-PPR demonstrates superior or highly competitive performance across all densities, effectively utilizing the abundance of paths in denser graphs.

\begin{figure}[h!]
    \centering
    \includegraphics[width=0.8\textwidth]{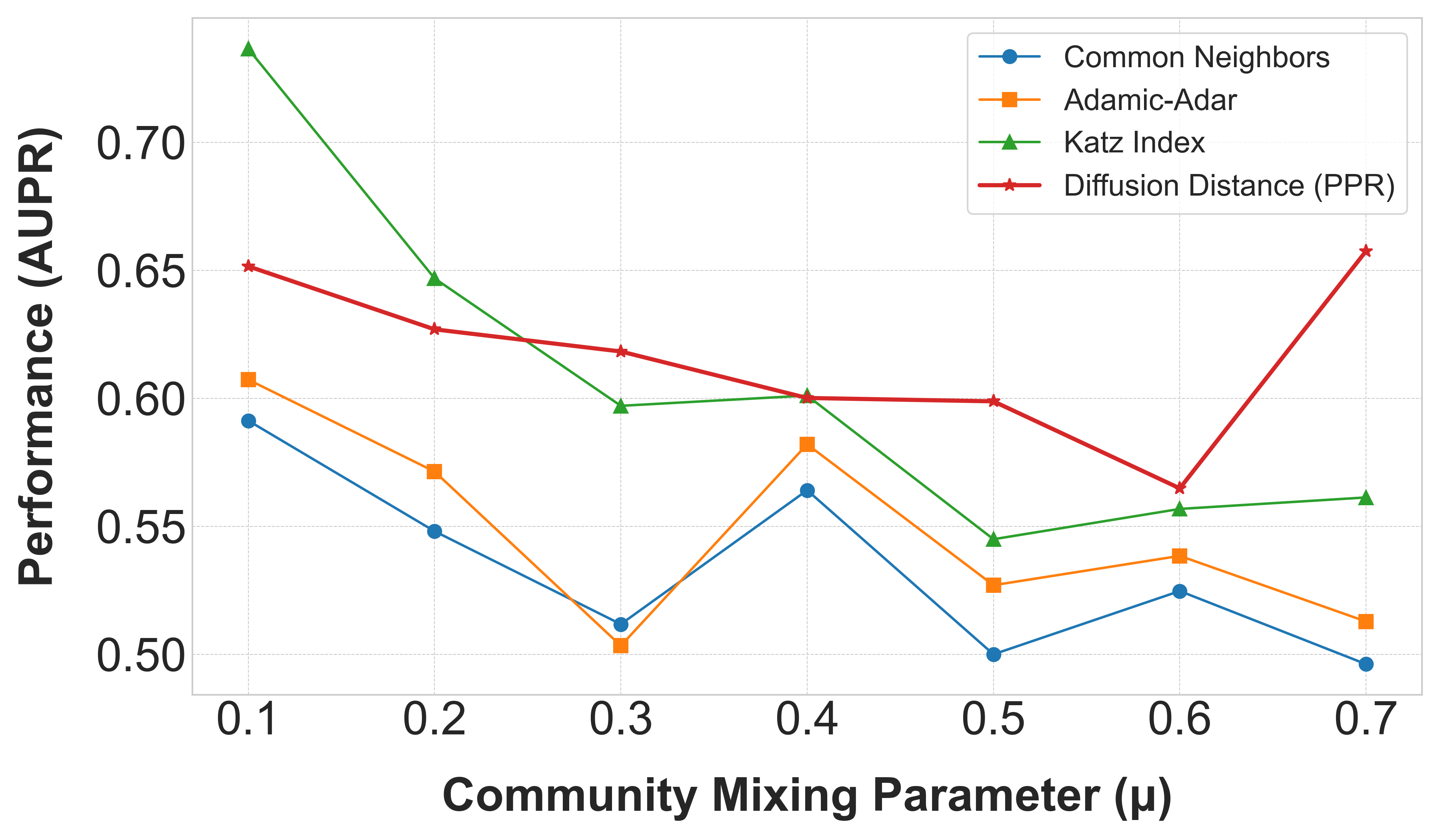}
    \caption{Robustness of link prediction methods against varying community structure in LFR benchmark networks. This figure plots the AUPR of each method as the community mixing parameter $\mu$ increases, where a higher $\mu$ signifies weaker and less distinct community structures. Networks were generated using the LFR benchmark model with $N=250$ nodes. While the performance of all methods declines as community structure weakens, the global heuristics (D-PPR and Katz Index) exhibit greater resilience compared to local methods, highlighting D-PPR's ability to identify structural similarities even when strong modularity is absent.}
    \label{fig:lfr_community}
\end{figure}

Effect of Community Structure: Figure \ref{fig:lfr_community} shows results on LFR networks. While all methods suffer as community structure weakens (increasing $\mu$), D-PPR and Katz exhibit greater resilience. D-PPR's robust performance highlights its ability to identify structural similarities even when modularity is low.

\subsection{Validation on Real-World Networks}

We first validate our framework on the well-known Zachary's Karate Club network\cite{zachary1977information}, a classic benchmark for community-related network tasks. As shown in Figure~4, the D-PPR method achieves a significantly higher AUPR score (0.800) compared to all baseline methods, including the strong global heuristic, the Katz Index (0.748). This result is particularly noteworthy because the Karate Club network's structure is largely governed by two distinct social factions. The superior performance of D-PPR suggests that its underlying mechanism---measuring the similarity of diffused influence fields (PPR vectors)---is highly effective at capturing this fundamental structural property. Nodes within the same faction have highly similar influence patterns, resulting in a small diffusion distance and thus a high link prediction score.

Moving to larger, more complex systems, our results on four large-scale empirical networks, shown in Figure~5, further confirm the effectiveness and robustness of the D-PPR framework. In all these networks, D-PPR significantly outperforms the local heuristics. Notably, it achieves the highest AUPR on the Email and WWW networks, which are known for their complex small-world and navigable properties. This demonstrates the framework's power and applicability to diverse real-world systems.

\begin{figure}[h!]
    \centering
    \includegraphics[width=0.8\textwidth]{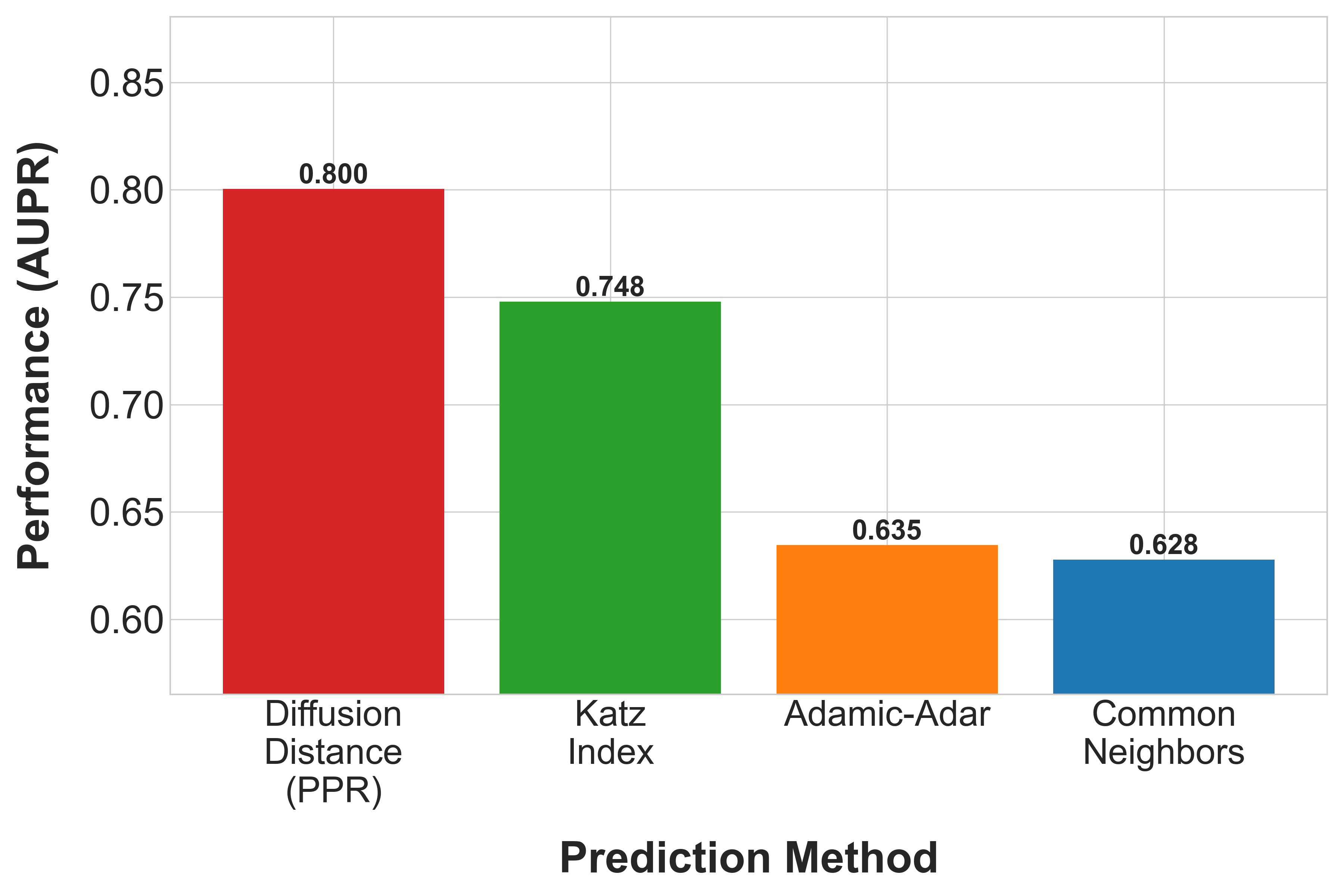}
    \caption{Comparative performance of link prediction methods on Zachary's Karate Club network. The bar chart displays the AUPR score for the proposed D-PPR framework against three baseline methods. The evaluation was performed on this classic social network benchmark by randomly hiding 20\% of the edges for testing. The results demonstrate that D-PPR significantly outperforms all other heuristics. This superior performance is attributed to its ability to effectively capture the network's well-defined community structure, which is the primary driver of its topology.}
    \label{fig:karate_bar}
\end{figure}


\begin{figure}[h!]
    \centering
    \includegraphics[width=0.95\textwidth]{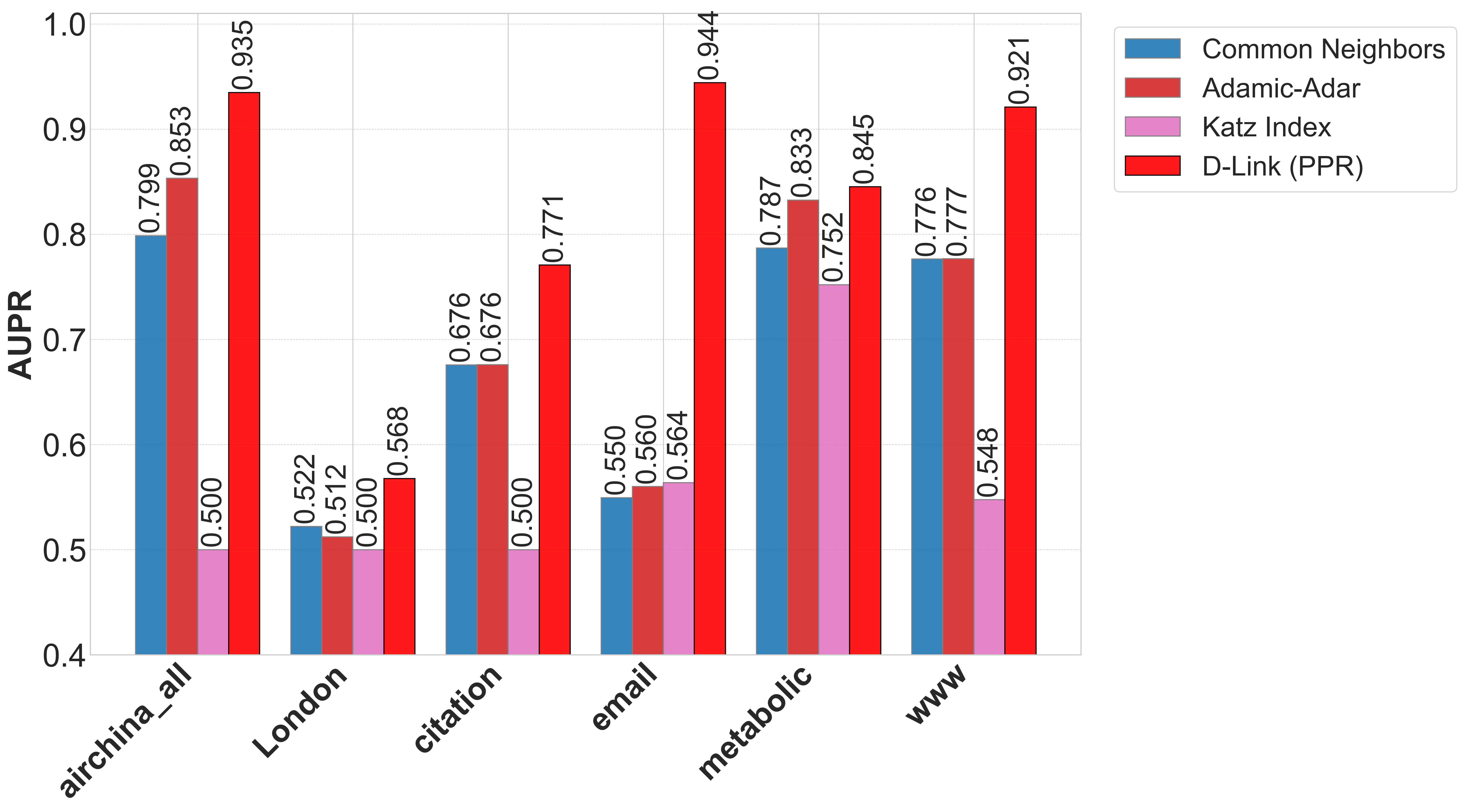}
    \caption{Performance evaluation on diverse large-scale real-world networks. This figure presents a comparative analysis of AUPR scores for D-PPR and baseline methods across seven real-world networks from technological, communication, biological, and transportation domains (see Table~\ref{tab:datasets} for network details). The results confirm the robust and broadly applicable performance of the D-PPR framework. Notably, it achieves the highest AUPR on several complex networks, such as Email and WWW, demonstrating its effectiveness and generalizability to diverse real-world systems.}
    \label{fig:real_world_bars}
\end{figure}
\section{Discussion and Conclusion}
In this work, we have introduced a novel theoretical framework for link prediction based on the physical process of diffusion. By conceptualizing a node's structural role as a Personalized PageRank signal and measuring the similarity of these signals through a diffusion metric, our D-PPR method offers a principled and powerful alternative to traditional static measures. The superior performance of D-PPR across a wide range of network models and real-world systems suggests that the dynamical processes a network can support are deeply informative of its underlying structure.

The physical intuition behind D-PPR is compelling: it measures a form of ``structural equivalence" based on how perturbations (represented by PPR signals) propagate from different nodes. If the diffused effects of two nodes are indistinguishable, they occupy similar topological positions, and a link between them is likely. This perspective connects the link prediction problem to broader concepts in network science, such as network geometry and spectral graph theory.

While we have demonstrated the efficacy of this framework, several avenues for future research are open. The extension of this dynamical approach to directed, weighted, and temporal networks is a natural next step. Furthermore, exploring the relationship between D-PPR and the architectures of Graph Neural Networks could inspire new, physics-informed models for graph representation learning.

In conclusion, we have shown that by unifying static topology with a dynamic diffusion process, we can construct a highly effective and universal framework for link prediction. This work not only provides a new state-of-the-art tool but also deepens our understanding of the intricate relationship between structure and dynamics in complex networks.

\section*{code availability}
Network data and code for replication and reuse have been deposited in \href{https://github.com/wanghuilin8/more}{Github} and \url{https://networksciencebook.com/translations/en/resources/data.html}.

\section*{Acknowledgements}
This work is supported in part by the National Key Research and Development Program of China under Grant No. 2024YFA1611003, the Fundamental Research Funds for the Central Universities (Grant No. CCNU25JC027), the 111 Project 2.0, with Grant No. BP0820038, the Anhui Provincial Department of Education(Grant No.2023AH050708).

\bibliographystyle{elsarticle-num} 
\bibliography{ppr}


\end{document}